\documentclass[conference]{IEEEtran}
\IEEEoverridecommandlockouts
\usepackage{cite}
\usepackage{amsmath,amssymb,amsfonts}
\usepackage{algorithmic}
\usepackage{graphicx}
\usepackage{textcomp}
\usepackage{xcolor}
\usepackage{url} % Useful for inserting web links nicely
\usepackage{array}
\usepackage{hyperref}

\definecolor{lightgreen}{rgb}{0.0, 0.5, 0.0}
\definecolor{lightred}{rgb}{0.8, 0.0, 0.0}

\def\BibTeX{{\rm B\kern-.05em{\sc i\kern-.025em b}\kern-.08em
    T\kern-.1667em\lower.7ex\hbox{E}\kern-.125emX}}
\begin{document}

\title{Multimodal Learning with Augmentation Techniques for Natural Disaster Assessment\\
\thanks{This work is supported by the Romanian Hub for Artificial Intelligence - HRIA, Smart Growth, Digitization and Financial Instruments Program, 2021-2027, MySMIS no. 334906.}

%\author{
%\IEEEauthorblockN{
%Adrian-Dinu Urse\IEEEauthorrefmark{1}, Dumitru-Clementin %Cercel\IEEEauthorrefmark{1}\IEEEauthorrefmark{2}, and Florin Pop\IEEEauthorrefmark{1}}
%\IEEEauthorblockA{
%\IEEEauthorrefmark{1}National University of Science and Technology POLITEHNICA Bucharest, Romania\\
%\ adrian\_dinu.urse@stud.acs.upb.ro, dumitru.cercel@upb.ro, florin.pop@upb.ro
% \IEEEauthorrefmark{3}
%}

\author{\IEEEauthorblockN{1\textsuperscript{st} Adrian-Dinu Urse}
\IEEEauthorblockA{\textit{NUST POLITEHNICA Bucharest} \\
Bucharest, Romania  \\
}
\and
\IEEEauthorblockN{2\textsuperscript{nd} Dumitru-Clementin Cercel\IEEEauthorrefmark{2}}
\IEEEauthorblockA{\textit{NUST POLITEHNICA Bucharest} \\
Bucharest, Romania  \\
}
\and
\IEEEauthorblockN{3\textsuperscript{rd} Florin Pop}
\IEEEauthorblockA{\textit{NUST POLITEHNICA Bucharest}\\
Bucharest, Romania  \\
}}

\thanks{\IEEEauthorrefmark{2} Corresponding author: dumitru.cercel@upb.ro.}

}

\maketitle

\begin{abstract}

Natural disaster assessment relies on accurate and rapid access to information, with social media emerging as a valuable real-time source. However, existing datasets suffer from class imbalance and limited samples, making effective model development a challenging task. This paper explores augmentation techniques to address these issues on the CrisisMMD multimodal dataset. For visual data, we apply diffusion-based methods, namely Real Guidance and DiffuseMix. For text data, we explore back-translation, paraphrasing with transformers, and image caption-based augmentation. We evaluated these across unimodal, multimodal, and multi-view learning setups. Results show that selected augmentations improve classification performance, particularly for underrepresented classes, while multi-view learning introduces potential but requires further refinement. This study highlights effective augmentation strategies for building more robust disaster assessment systems.

\end{abstract}

\begin{IEEEkeywords}
Data augmentation, diffusion models, multimodal learning.
\end{IEEEkeywords}

\section{Introduction}

Natural disaster assessment is a critical application area where the identification of timely and accurate information can significantly improve humanitarian response efforts. The use of social networks has gained attention as a complementary source for disaster analysis, offering real-time information from affected areas. However, leveraging such data effectively remains challenging due to the inherent limitations of existing datasets, including class imbalance, limited sample sizes, as well as noisy and unstructured content.

To overcome these challenges, this research investigates advanced data augmentation techniques that aim to enhance dataset diversity and improve model performance for natural disaster classification. Using the multimodal CrisisMMD \cite{crisismmd} dataset, which comprises both textual and visual information from disaster-related tweets, this work applies targeted augmentation strategies for both modalities.

Beyond improving the unimodal performance, this paper also examines how these augmentation techniques affect multimodal and multi-view setups, which combine text and image information.

The following is a summary of this study's primary contributions:
\begin{itemize}
    \item We evaluate the effectiveness of diffusion-based image augmentation techniques, namely Real Guidance  \cite{real-gui} and DiffuseMix \cite{diffusemix}, enhancing classification performance for disaster-related visual data.
    \item We analyze various text augmentation methods, including back-translation, paraphrasing using transformers \cite{vaswani2017attention}, and image-caption-based augmentation, and assess their effects on disaster classification.
    \item We explore multimodal learning configurations that utilize both modalities, demonstrating their superiority over unimodal approaches to disaster assessment.
    \item We evaluated a multi-view learning methodology that integrates both original and augmented data representations during training, highlighting its potential and limitations for disaster assessment tasks.
\end{itemize}

\section{Related Work}
Disaster assessment means quickly identifying key information from real-time data sources, including social media and satellite images. To support faster decision-making, researchers have explored different ways to automate this process.

Early research in this field concentrated on analyzing the textual content of social media posts, especially those from Twitter. Classic machine learning models, such as Support Vector Machines, Logistic Regression, and Random Forest, were commonly used to classify posts as informative or non-informative and to further assign them to humanitarian categories, including infrastructure damage and rescue efforts \cite{survey}. Imran et al. \cite{aidr} introduced the Artificial Intelligence for Disaster Response platform, which used machine learning classifiers and crowd-sourced annotations to classify tweets during the 2013 earthquake in Pakistan. Another significant early work was CrisisLex \cite{crisislex}, introduced by Olteanu et al., which provided a lexicon of crisis-related words, used to filter and identify crisis-related tweets.

Caragea et al. \cite{naive-b-crisis} addressed the issue of limited data in crisis scenarios by applying domain adaptation methods. They trained a Naive Bayes model on past disasters and adapted it to use in unseen crisis situations. The introduction of transformer-based language models has considerably progressed the field, providing stronger context understanding and improved classification performance. Pre-trained models, such as BERT \cite{bert}, RoBERTa \cite{roberta}, and DeBERTa \cite{deberta}, have been fine-tuned on crisis-specific datasets, including CrisisLexT26, CrisiNLP, and CrisisMMD, resulting in more accurate outcomes. Liu et al. \cite{CrisisBERT} introduced CrisisBERT, a distilled version of BERT tailored for classifying crisis-related tweets, and showed that it achieved notable gains across multiple tasks. In the same way, Lamsal et al. \cite{CrisisTransformers} developed CrisisTransformers, a set of transformer-based models fine-tuned on a large-scale corpus of over 15 billion tokens from tweets spanning more than 30 crisis events.

Although early work focused on text, images shared during disasters of damaged infrastructure have also proven valuable. Alam et al. \cite{alam} introduced benchmarks for disaster-related image classification, which include tasks such as identifying disaster types, evaluating information, determining humanitarian relevance, and estimating damage severity. 

Using convolutional neural networks, including VGG, ResNet, AlexNet, and InceptionNet, they showed that image-based models can perform well even without accompanying text. Transformer-based models have also been used to assess disasters, just as with text. CrisisViT \cite{CrisisViT} is a vision transformer designed for classifying crisis-related images, outperforming both convolutional baselines and fine-tuned ViT models across various tasks, such as emergency category classification, damage severity assessment, and humanitarian relevance. The model used a two-stage training process, initially using the extensive Incidents1M dataset \cite{incidents1m} for pre-training, followed by fine-tuning on conventional standard benchmarks.

Multimodal learning has gained interest, as combining text and images offers a fuller understanding of crisis situations than using either modality alone. One of the first studies in this direction was conducted by Ofli et al. \cite{ofli}, which showed that integrating visual and textual information from CrisisMMD through early fusion significantly improved classification performance compared to unimodal models. Then, Abavisani et al. \cite{abavisani} proposed a stronger architecture, combining a pre-trained BERT model for text and a DenseNet \cite{densenet} model for images using a cross-attention fusion mechanism. Their approach outperformed both unimodal baselines and simpler multimodal models. Further advancing this direction, Gupta et al. \cite{crisiskan} introduced CrisisKan, a knowledge-augmented multimodal model that incorporated external information from Wikipedia. Furthermore, they involved a guided cross-attention mechanism to fuse text and image features more effectively. Their model achieved superior performance across all CrisisMMD classification tasks. 

%Teng et al. \cite{crisis-ambi} addressed a novel problem, namely the issue of noisy and ambiguous social media content. Thus, they filtered CrisisMMD to keep only examples where the text and image labels were consistent. They developed a multimodal model that combined BERT and VGG16 through cross-attention fusion, producing stronger results, particularly in ambiguous cases, and highlighting the advantages of multimodal approaches in real-world disaster scenarios.

\section{Dataset}

The CrisisMMD dataset \cite{crisismmd} is a multimodal collection of Twitter posts from major disasters in 2017, including text and accompanying images that capture real-time moments during crises.

In this work, we focus on the humanitarian classification task,
i.e., classifying tweets into specific humanitarian categories. The goal is to capture the type and urgency of the information shared during disasters. The categories are as follows: 
\begin{itemize}
    \item Affected Individuals - posts about people that are impacted, injured, missing, or in need of help;
    \item  Infrastructure and Utility Damage - reports of damage to vehicles, buildings, roads, bridges, and power lines;
    \item Not Humanitarian - posts that are unrelated to humanitarian needs or disaster relief;
    \item Other relevant Information - important information, useful for humanitarian response, that does not fall into the other categories;
    \item Rescue, Volunteering, or Donation Effort - posts highlighting ongoing rescue operations, volunteering, or donation efforts.
\end{itemize}

For our experiments, we used the agreed-upon annotation subset of CrisisMMD, which includes only posts where annotators assigned the same label. The dataset is divided into three subsets as follows: a training set of 6,126 samples, a validation set of 998 samples, and a test set of 955 samples. We note that the distribution of the classes is imbalanced, as shown in Table~\ref {tab:class_distribution}, with the "Not Humanitarian" class being the most common, while other classes, such as "Affected Individuals," have fewer samples. The imbalanced distribution is a challenge for the models to perform well, which motivates the application of the augmentation techniques explored in this study.

\begin{table}[h]
\centering
\begin{tabular}{|l|c|c|c|}
\hline
\textbf{Class Label} & \textbf{Train} & \textbf{Dev} & \textbf{Test} \\
\hline
Affected Iindividuals & 71   & 9   & 9   \\
\hline
Infrastructure and Utility Damage & 612  & 80  & 81  \\
\hline
Not Humanitarian & 3,252 & 521 & 504 \\
\hline
Other Relevant Information & 1,279 & 239 & 235 \\
\hline
Rescue, Volunteering, or Donation Effort & 912  & 149 & 126 \\
\hline
\textbf{Total} & \textbf{6,126} & \textbf{998} & \textbf{955} \\
\hline
\end{tabular}
\caption{Class distribution of the dataset across training, Validation, and test splits.}
\label{tab:class_distribution}
\end{table}

\section{Methodology}
This study investigates two problems: the impact of data augmentation techniques as well as multimodal learning approaches on enhancing disaster assessment using the CrisisMMD dataset. Our experiments focus on tackling class imbalance and increasing data diversity through text and image augmentation, while also evaluating the impact of these techniques in unimodal and multimodal classification conditions.

\subsection{Data Augmentation for Images}
We explore two diffusion-based image augmentation methods to enhance the visual diversity of the CrisisMMD dataset and mitigate class imbalance.

The first method is Real Guidance \cite{real-gui}, a conditioned image-to-image generation method, with the Stable Diffusion 1.5 model. Each original image in the training set was slightly modified using a combination of a reference image and a text prompt, resulting in realistic synthetic images that kept the disaster-related features. This process doubled the size of the training dataset while maintaining contextual relevance.

The second approach, DiffuseMix \cite{diffusemix}, is an advanced augmentation technique recently introduced in the literature. DiffuseMix uses prompt-based transformations, masked blending, and fractal-based visual modifications to create diverse augmented images. To guide the generation process in introducing stylistic differences while preserving the original images' structural content, we employ predefined prompts such as "autumn", "watercolor art", "sunset", or "mosaic". To further enhance visual complexity, we applied fractal blending during the augmentation.

We only utilize DiffuseMix augmentation on classes that were underrepresented in the dataset, namely "Affected Individuals", "Infrastructure and Utility Damage", and "Rescue, Volunteering, or Donation Effort". This targeted approach increased the number of samples for these classes, thereby helping to balance the dataset and improving the model's performance on minority categories.

To evaluate the impact of both augmentation techniques, several image classification models were fine-tuned and tested, including convolutional architectures (such as ResNet18 and ResNet50) and transformer-based models (such as ViT and MambaViT \cite{mambavit}). 

\subsection{Data Augmentation for Text}
To increase linguistic diversity and improve model generalization for disaster assessment, three text augmentation strategies were employed on the CrisisMMD dataset.

The first method used back-translation, a widely adopted technique that introduces subtle linguistic variations while preserving the original meaning. We apply a multistep translation process, in which English tweets are first translated into French, then into German, back into French, and finally returned to English. This process created paraphrased versions of the original tweets, providing additional training examples and maintaining semantic consistency.

The second approach involved paraphrasing with a transformer-based language model. Using the open-source Mistral-7B-Instruct model\footnote{\url{https://huggingface.co/mistralai/Mistral-7B-Instruct-v0.2}}, we rewrote the tweets to produce diverse versions while preserving their meaning and Twitter-specific style, including hashtags, emojis, and informal language. To ensure the quality of the generated text, we used a filtering process, retaining only those that were label-consistent, semantically diverse, and within acceptable length ranges.

The third augmentation strategy was caption-based augmentation, which used information from the visual modality. For each image, we generate a descriptive caption using the BLIP-2 \cite{blip} model, a vision-language architecture designed for image captioning. Then, we concatenated the generated descriptions with the corresponding tweet text, enriching the textual input with an additional image-derived context.

To assess the efficacy of these augmentation techniques, the best text classification models identified during baseline experiments were fine-tuned and evaluated on the augmented datasets.

\subsection{Multimodal and Multi-View Learning}

This work extends the unimodal experiments by exploring multimodal classification through the integration of textual and visual information to improve the accuracy of disaster assessment. We conducted experiments on both the original CrisisMMD dataset and its augmented versions, combining text augmentations with image augmentations. Specifically, back-translated textual data was paired with both Real Guidance and DiffuseMix image augmentations.

An early fusion strategy was adopted to integrate the textual and visual features. Various image encoders, including convolutional and transformer-based architectures, were combined with pre-trained text encoders. This setup enables the model to leverage complementary information from both modalities for more accurate predictions.

To further enrich the model's capacity to capture diverse aspects of the data, we explore a multi-view learning approach. In this setting, additional augmented views of both texts and images, as well as image-generated captions produced by BLIP-2, were incorporated within each training instance. Unlike conventional augmentation strategies, where augmented pairs are treated as separate training examples, multi-view learning preserves all representations (i.e., original text, original image, augmented text, augmented image, and caption) within a unified input structure.

\begin{figure*}[ht]
\centering
\includegraphics[width=0.6\textwidth, height=0.29\textheight]{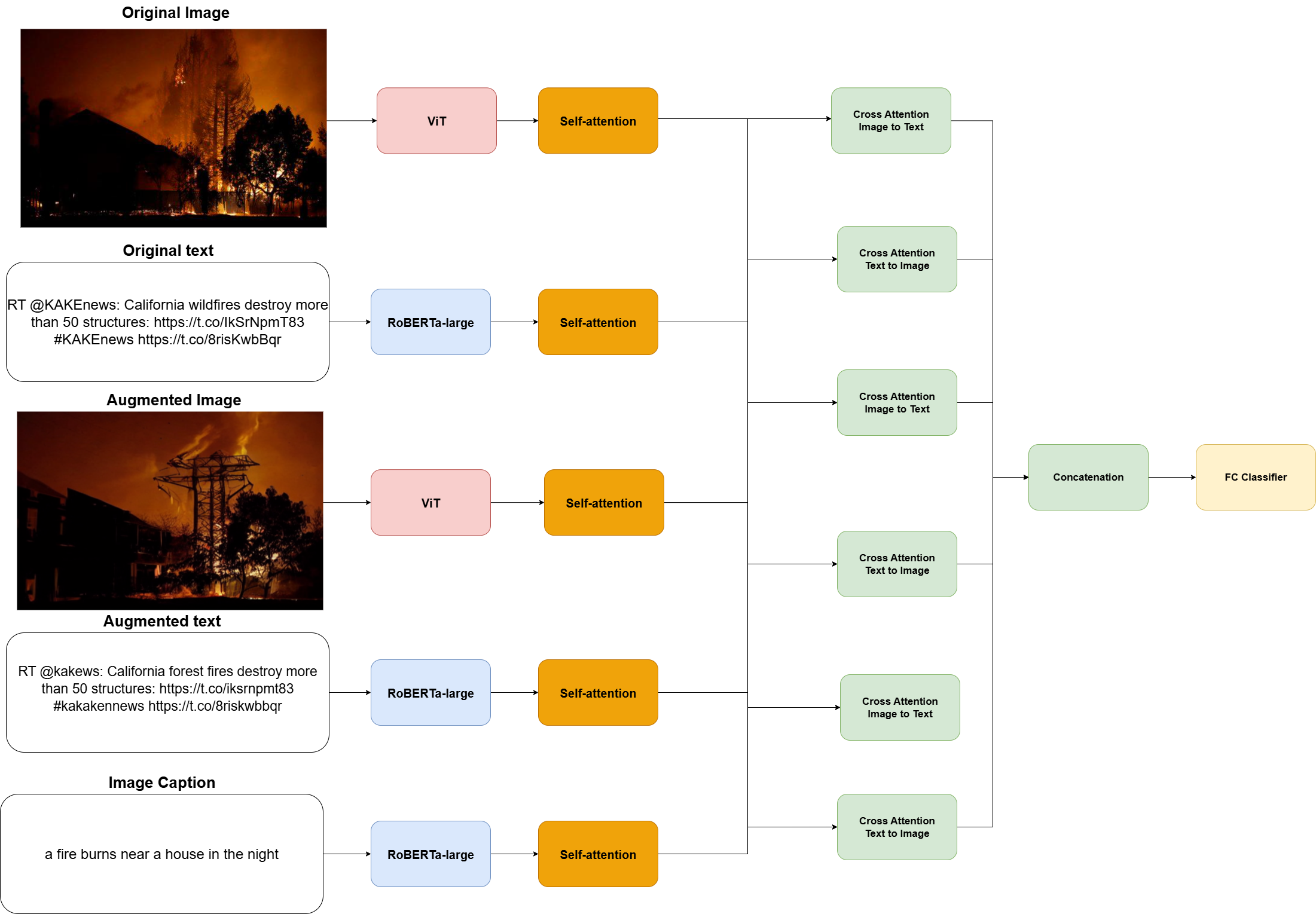}
\caption{Our multi-view learning architecture.}
\label{fig:multi_view}
\end{figure*}

During testing, only the original text-image pairs are provided, while the additional views serve as empty placeholders. This strategy allows the model to benefit from the enriched training signal introduced by the augmented views and captions, while relying solely on the original inputs during evaluation.

Several architectural configurations were tested to effectively combine the different views, including models with cross-attention mechanisms designed to facilitate deeper interaction between modalities and augmented views, as seen in Figure \ref{fig:multi_view}. Additionally, self-attention layers were introduced before cross-attention to enhance intra-modal feature extraction, thereby improving cross-modal alignment.

\section{Experimental Setup}

%\subsection{Dataset Splits}
%The experiments were conducted on the agreed annotation subset of the CrisisMMD dataset, which includes only samples where annotators consistently assigned the same label. The dataset was divided into a training set of 6,126 samples, a validation set of 998 samples, and a test set of 955 samples.

\subsection{Unimodal Classification}

For image classification, all images were resized to $224 \times 224$ pixels. We employed pre-trained models and fine-tuned them for 10 epochs on both the original dataset and its augmented versions. We used the AdamW optimizer for all experiments. 

For text classification, we tokenized the inputs using the specific tokenizer associated with each pre-trained language model. All text models were also fine-tuned for 10 epochs using the AdamW optimizer, just like the image pipeline.  

Classification experiments were first performed on the original dataset, thereby setting the baseline results. Then, the best models for both text and image classification were chosen and tested on the augmented datasets to assess the impact of the proposed augmentation techniques.

\subsection{Multimodal and Multi-View Classification}

For multimodal classification, an early fusion strategy was adopted, in which textual and visual characteristics were concatenated and used for the final prediction. The same training setup as in the unimodal experiments was applied, including the use of the AdamW optimizer and 10 training epochs.

The multi-view learning experiments followed the same early fusion structure, with additional augmented views incorporated during the training process.

\subsection{Evaluation Metrics}

We evaluated the model performance using accuracy (Acc) and weighted F1-score. The weighted F1-score was selected to account for the imbalanced class distribution within the dataset, providing a more reliable measure of overall classification performance.

In addition to model performance, quantitative metrics were also calculated to assess the quality of the augmented textual data. These metrics evaluate both the semantic preservation and textual similarity of the augmented tweets with respect to the originals:

\begin{itemize}
    \item \textbf{Semantic Similarity} measures how closely the meaning of the augmented tweet aligns with the original. Sentence embeddings were generated using the \textit{sentence-transformers/all-mpnet-base-v2} model\footnote{\url{https://huggingface.co/sentence-transformers/all-mpnet-base-v2}}, and cosine similarity was calculated between the original and augmented representations.
    
    \item \textbf{ROUGE-L} captures the degree of textual overlap based on the longest common subsequence, providing insight into how much of the original wording is retained in the augmented version.
    
    \item \textbf{Perplexity} estimates the fluency of the augmented tweets. A pre-trained GPT-2 language model\footnote{\url{https://huggingface.co/openai-community/gpt2}} was used to calculate perplexity, with lower scores indicating more fluent and predictable text.
\end{itemize}

\section{Results and Discussion}

\subsection{Impact of Image Augmentation}

\subsubsection{Real Guidance}

The synthetic images generated through the Real Guidance-conditioned image-to-image transformation process maintained high visual quality, preserving the essential elements of the original disaster images while introducing subtle yet meaningful variations. The augmented images presented minimal signs of hallucinations or artifacts, ensuring that visual diversity was increased without significantly distorting the original context.

\begin{table*}[ht]
\centering
\caption{Performance (Accuracy, F1-score) of models using DiffuseMix on a single class.}
\renewcommand{\arraystretch}{1.25}
\setlength{\tabcolsep}{4pt}
\begin{tabular}{|c|c|c|c|c|c|c|c|c|}
\hline
\textbf{Model} 
& \multicolumn{2}{c|}{\textbf{Original}} 
& \multicolumn{2}{c|}{\textbf{Affected-Ind}} 
& \multicolumn{2}{c|}{\textbf{Infrastructure}} 
& \multicolumn{2}{c|}{\textbf{Rescue}} \\
\cline{2-9}
& \textbf{Acc} & \textbf{F1} & \textbf{Acc} & \textbf{F1} & \textbf{Acc} & \textbf{F1} & \textbf{Acc} & \textbf{F1} \\
\hline
ResNet18            & 0.7634 & 0.7584 & 0.7717 & 0.7655 & 0.7780 & 0.7759 & 0.7696 & 0.7663 \\
\hline
ResNet50            & 0.7654 & 0.7601 & 0.7675 & 0.7619 & 0.7759 & 0.7709 & 0.7717 & 0.7646 \\
\hline
ViT                 & 0.8106 & 0.8069 & 0.8046 & 0.8016 & 0.8126 & 0.8092 & 0.8066 & 0.8036 \\
\hline
ViT-frozen          & --     & --     & 0.7826 & 0.7745 & 0.7717 & 0.7687 & 0.7906 & 0.7849 \\
\hline
ViT-frozen-partial  & --     & --     & \textbf{0.8156} & \textbf{0.8155} & 0.8216 & 0.8189 & \textbf{0.8226} & \textbf{0.8210} \\
\hline
MambaViT            & \textbf{0.8206} & \textbf{0.8190} & 0.8026 & 0.8013 & \textbf{0.8246} & \textbf{0.8229} & 0.8066 & 0.8022 \\
\hline
\end{tabular}
\label{tab:diff_sing_class}
\end{table*}

\begin{table*}[ht]
\centering
\caption{Performance (Accuracy, F1-score) of models using DiffuseMix with 2-3 class combinations.}
\renewcommand{\arraystretch}{1.25}
\setlength{\tabcolsep}{4pt}
\begin{tabular}{|c|c|c|c|c|c|c|}
\hline
\textbf{Model} 
& \multicolumn{2}{c|}{\textbf{Original}} 
& \multicolumn{2}{c|}{\textbf{Affected-Infra}} 
& \multicolumn{2}{c|}{\textbf{Affected-Infra-Rescue}} \\
\cline{2-7}
& \textbf{Acc} & \textbf{F1} & \textbf{Acc} & \textbf{F1} & \textbf{Acc} & \textbf{F1} \\
\hline
ResNet18           & 0.7634 & 0.7584 & 0.7665 & 0.7492 & 0.7675 & 0.7601 \\
\hline
ResNet50           & 0.7654 & 0.7601 & 0.7696 & 0.7656 & 0.7771 & 0.7671 \\
\hline
ViT                & 0.8106 & 0.8069 & 0.8026 & 0.8031 & 0.8126 & 0.8120 \\
\hline
ViT-frozen         & --     & --     & --     & --     & 0.7846 & 0.7845 \\
\hline
ViT-frozen-partial & --     & --     & --     & --     & \textbf{0.8186} & \textbf{0.8164} \\
\hline
MambaViT           & \textbf{0.8206} & \textbf{0.8190} & \textbf{0.8126} & \textbf{0.8077} & 0.8116 & 0.8118 \\
\hline
\end{tabular}
\label{tab:diff_two_class}
\end{table*}

Based on the initial classification experiments conducted on the original dataset, the best models (i.e., ViT, EfficientNet\_b0, and MambaViT) were selected for further testing on the augmented dataset. Two convolutional models, ResNet18 and ResNet50, were also used to see how augmentation affects different types of architectures.

\begin{table}[ht]
\centering
\caption{Classification results using Real Guidance augmentation.}
\renewcommand{\arraystretch}{1.3}
\resizebox{\columnwidth}{!}{
\begin{tabular}{|>{\centering\arraybackslash}p{2.0cm}|>{\centering\arraybackslash}p{1.0cm}|>{\centering\arraybackslash}p{1.5cm}|>{\centering\arraybackslash}p{1.2cm}|>{\centering\arraybackslash}p{1.2cm}|}
\hline
\textbf{Model} & \multicolumn{2}{c|}{\textbf{Original}} & \multicolumn{2}{c|}{\textbf{Real Guidance Aug.}} \\
\cline{2-5}
 & \textbf{Acc} & \textbf{F1-score} & \textbf{Acc} & \textbf{F1-score} \\
\hline
VGG19 & 0.7634 & 0.7550 & -- & -- \\
\hline
ResNet18 & 0.7634 & 0.7584 & 0.7717 & 0.7639 \\
\hline
ResNet50 & 0.7654 & 0.7601 & 0.7717 & 0.7605 \\
\hline
ResNet101 & 0.7666 & 0.7533 & -- & -- \\
\hline
EfficientNet\_b0 & 0.8105 & 0.8043 & 0.7958 & 0.7919 \\
\hline
ViT & 0.8106 & 0.8069 & \textbf{0.8066} & \textbf{0.8008} \\
\hline
MambaViT & \textbf{0.8206} & \textbf{0.8190} & 0.8056 & 0.7967 \\
\hline
Swin Transformer & 0.7936 & 0.7914 & -- & -- \\
\hline
\end{tabular}}
\label{tab:real_gui_results}

\end{table}

The results, summarized in Table~\ref{tab:real_gui_results}, showed clear differences in the way different models responded to the augmented dataset. The accuracy and F1-scores for ResNet18 and ResNet50 both increased regularly, indicating that these convolutional architectures effectively utilized the extra visual diversity introduced by Real Guidance. However, the performance of EfficientNet\_b0 dropped, suggesting challenges in adapting to the augmented variations.

For transformer-based models, including ViT and MambaViT, a decrease in both accuracy and F1-score was observed on the augmented dataset. These results indicate that the augmentations may have introduced visual noise, which interfered with the attention mechanisms of these models, causing them to focus on irrelevant artifacts rather than the essential features of the disaster scenes. These findings suggest that convolutional networks demonstrated greater robustness to noise induced by augmentation in this context.

\subsubsection{DiffuseMix}

The images generated with DiffuseMix had clear stylistic transformations and fractal-based textures as intended. The fractal patterns were visually diverse, yet they sometimes seemed strange and unnatural, adding unusual visual elements to the dataset.

The classification results on the datasets augmented with DiffuseMix, reported in Tables \ref{tab:diff_sing_class} and \ref{tab:diff_two_class}, presented patterns comparable to those identified with Real Guidance. The convolutional models, ResNet18 and ResNet50, consistently benefited from the augmentations, achieving greater accuracy and F1-scores for most class combinations. This shows even more how convolutional architectures can use increased visual diversity to comprehend visual features.

Transformer-based models had mixed behavior. ViT demonstrated improved performance only when the "Infrastructure and Utility Damage" class was augmented and when all three underrepresented classes were augmented. MambaViT showed a performance increase only when "Infrastructure and Utility Damage" was augmented individually.

Additionally, experiments with frozen ViT variants revealed that the fully frozen model always underperformed on the augmented datasets, indicating that it cannot adapt to the new visual styles introduced by DiffuseMix. However, the partially frozen ViT model, where only the final layers were trainable, performed better across all augmentation scenarios, highlighting the benefits of controlled fine-tuning in balancing feature stability with adaptability.

\subsection{Impact of Text Augmentation}

\subsubsection{Back-translation}

The back-translation method worked well in incorporating minor linguistic differences while preserving the original meaning. An example is presented below:

\begin{itemize}
    \item \textbf{Original text:} \textit{Emergency medical supplies delivered tonight to Sonoma Public Health for evacuees of Northern California Wildfires.}
    \item \textbf{Text after back-translation:} \textit{Emergency medical care this evening at Sonoma Public Health for the evacuation of Nordkalifornia Wildfires.}
\end{itemize}

Following baseline experiments on the original dataset, RoBERTa-large, DeBERTa-v3-large, and BERTweet-base emerged as the top-performing models, with RoBERTa-large achieving the highest accuracy of 84.29\% and an F1-score of 84.12\%. All three models were subsequently fine-tuned and evaluated on the augmented dataset. In all models, a slight but consistent improvement in both accuracy and F1-score was achieved, as presented in Table \ref{tab:unimodal_backtr}, indicating that the back-translation effectively improved linguistic diversity without compromising the consistency of the label.

\begin{table}[ht]
\centering
\caption{Performance (Accuracy, F1-score) of various models with and without back-translation augmentation.}
\renewcommand{\arraystretch}{1.3}
\resizebox{\columnwidth}{!}{
\begin{tabular}{|>{\centering\arraybackslash}p{2.0cm}|>{\centering\arraybackslash}p{1.0cm}|>{\centering\arraybackslash}p{1.5cm}|>{\centering\arraybackslash}p{1.2cm}|>{\centering\arraybackslash}p{1.2cm}|}
\hline
\textbf{Model} & \multicolumn{2}{c|}{\textbf{Original}} & \multicolumn{2}{c|}{\textbf{Back-Translation Aug.}} \\
\cline{2-5}
 & \textbf{Acc} & \textbf{F1-score} & \textbf{Acc} & \textbf{F1-score} \\
\hline
BERT-base & 0.8042 & 0.8044 & - & - \\
\hline
DeBERTa-v3-base & 0.8220 & 0.8172 & - & - \\
\hline
DeBERTa-v3-large & 0.8346 & 0.8314 & 0.8377 & 0.8345 \\
\hline
RoBERTa-base & 0.8346 & 0.8324 & - & - \\
\hline
RoBERTa-large & \textbf{0.8429} & \textbf{0.8412} & \textbf{0.8471} & \textbf{0.8455} \\
\hline
mpnet-base-v2 & 0.8188 & 0.8144 & - & - \\
\hline
ELECTRA-base & 0.8262 & 0.8252 & - & - \\
\hline
BERTweet-base & 0.8262 & 0.8251 & 0.8346 & 0.8282 \\
\hline
\end{tabular}}
\label{tab:unimodal_backtr}
\end{table}

\subsubsection{Paraphrasing with Transformers}

The paraphrasing approach generated various variations in tweets while maintaining semantic integrity and adapting to the informal style typical of social media posts. An example of paraphrasing generated by the Mistral-7B-Instruct model is provided below:

\begin{itemize}
    \item \textbf{Original text:} \textit{Please help support and share!!! \#LasVegasShooting \#mexicoearthquake \#PuertoRicoSeLevanta.}
    \item \textbf{Paraphrased text:} \textit{URGENT Las Vegas shootings, Mexico quake, PR rising! Join forces, retweet \#LasVegasStrong \#Mexico.}
\end{itemize}

Table \ref{tab:unimodal_para} shows that all the models that were tested performed better after being trained on the dataset that included paraphrased tweets. BERTweet-base initially performed the lowest, but its performance improved significantly with the addition of more training examples. In general, this method of augmentation improved the model's ability to generalize to different tweet formulations while maintaining consistent labels.

\begin{table}[ht]
\centering
\caption{Performance comparison on CrisisMMD with and without paraphrasing-based augmentation.}
\label{tab:unimodal_para}
\renewcommand{\arraystretch}{1.3}
\resizebox{\columnwidth}{!}{
\begin{tabular}{|>{\centering\arraybackslash}p{2.0cm}|>{\centering\arraybackslash}p{1.0cm}|>{\centering\arraybackslash}p{1.5cm}|>{\centering\arraybackslash}p{1.2cm}|>{\centering\arraybackslash}p{1.2cm}|}
\hline
\textbf{Model} & \multicolumn{2}{c|}{\textbf{Original}} & \multicolumn{2}{c|}{\textbf{Paraphrase Aug.}} \\
\cline{2-5}
 & \textbf{Acc} & \textbf{F1-score} & \textbf{Acc} & \textbf{F1-score} \\
\hline
DeBERTa-v3-large & 0.8346 & 0.8314 & 0.8450 & 0.8465 \\
\hline
RoBERTa-large & \textbf{0.8429} & \textbf{0.8412} & \textbf{0.8461} & \textbf{0.8469} \\
\hline
BERTweet-base & 0.8262 & 0.8251 & 0.8429 & 0.8386 \\
\hline
\end{tabular}}
\end{table}

\subsubsection{Augmentation with Image Captioning}

Before adding the generated image captions to the textual input, their quality was manually checked. Most of the captions were correct and clearly described the most important parts of the corresponding disaster images. Below is an example of an augmented tweet:

\begin{itemize}
    \item \textbf{Original tweet with image caption:} \textit{\#smaili community steps up to serve \#Harvey victims. A woman standing at a table with boxes of food.}
\end{itemize}

\begin{table}[ht]
\centering
\caption{Performance comparison (accuracy, F1-score) on CrisisMMD with and without image caption-based text augmentation.}
\label{tab:unimodal_cap}
\renewcommand{\arraystretch}{1.3}
\resizebox{\columnwidth}{!}{
\begin{tabular}{|>{\centering\arraybackslash}p{2.0cm}|>{\centering\arraybackslash}p{1.0cm}|>{\centering\arraybackslash}p{1.5cm}|>{\centering\arraybackslash}p{1.2cm}|>{\centering\arraybackslash}p{1.2cm}|}
\hline
\textbf{Model} & \multicolumn{2}{c|}{\textbf{Original}} & \multicolumn{2}{c|}{\textbf{Caption-Based Aug.}} \\
\cline{2-5}
 & \textbf{Acc} & \textbf{F1-score} & \textbf{Acc} & \textbf{F1-score} \\
\hline
DeBERTa-v3-large & 0.8346 & 0.8314 & 0.8209 & 0.8187 \\
\hline
RoBERTa-large & \textbf{0.8429} & \textbf{0.8412} & \textbf{0.8387} & \textbf{0.8308} \\
\hline
BERTweet-base & 0.8262 & 0.8251 & 0.8126 & 0.8063 \\
\hline
\end{tabular}}
\end{table}

Although the image captions added contextual information, this augmentation consistently reduced performance, with accuracy and F1-scores dropping by approximately 1\% (see Table \ref{tab:unimodal_cap}). The likely reason is that the captions were only added during training, while the test set remained unchanged. This mismatch between the training and evaluation data introduced a distribution shift that could cause models to overfit to features that are present only in the augmented training
set. Consequently, the models struggled to generalize during
the evaluation.

\subsubsection{Quantitative Evaluation of Text Augmentation}

To complement the classification experiments, a quantitative evaluation of the augmented texts was performed using semantic similarity, ROUGE-L, and perplexity metrics. Table~\ref{tab:text_quality} presents the average scores for each augmentation method, calculated over the training set.

\begin{table}[h]
\centering
\caption{Quantitative evaluation of augmented texts.}
\setlength{\tabcolsep}{3pt}
\renewcommand{\arraystretch}{1.1}
\begin{tabular}{|p{2.5cm}|c|c|c|}
    \hline
    \textbf{Augmentation Method} & \textbf{Sem. Sim.} & \textbf{ROUGE-L} & \textbf{Perplexity} \\
    \hline
    Paraphrasing & 0.7893 & 0.3067 & 200.11\\
    \hline
    Back-translation & 0.9833 & 0.9618 & 134.85\\
    \hline
    Image caption & 0.8886 & 0.7942 & 115.12\\
    \hline
\end{tabular}
\label{tab:text_quality}
\end{table}

The original training set had an average perplexity of 110.15. This value is relatively high because tweet-style text is often informal and unstructured, with hashtags, URLs, abbreviations, and slang.

The back-translation achieved the highest semantic similarity and ROUGE-L scores among the augmentation methods, indicating that it best preserved both the meaning and textual structure of the original tweets. However, it also resulted in a slightly higher perplexity than the original, indicating increased linguistic complexity and reduced predictability.

The captions generated from the images provided a good balance, maintaining a moderate level of meaning and lexical overlap while achieving the lowest perplexity among all methods.

The paraphrasing introduced the greatest textual variation, as shown by the lowest semantic similarity and ROUGE-L scores and the highest perplexity. This result aligns with the fact that paraphrasing introduces more linguistic diversity, as it alters the wording and structure more than other augmentation approaches.

\subsection{Results for Multimodal Learning}

The results of the multimodal classification experiments show the differential effects of image and text augmentations across various model configurations (see Table~\ref{tab:tabel_9}). Three of the tested models—BERT-VGG, BERT-ViT, and RoBERTa-ViT—all got better when back-translated text data was combined with Real Guidance image augmentations. The RoBERTa-ViT model achieved the largest relative gains in both accuracy and F1-score, indicating that the additional information from the complementary augmentations was useful to the model.

\begin{table}[ht]
\centering
\caption{Performance (Accuracy, F1-score) of multi-modal models with RealGuidance and Back-translation augmentations.}
\label{tab:tabel_9}
\renewcommand{\arraystretch}{1.25}
\setlength{\tabcolsep}{3pt}
\begin{tabular}{|>{\centering\arraybackslash}p{2.0cm}|>{\centering\arraybackslash}p{1.0cm}|>{\centering\arraybackslash}p{1.5cm}|>{\centering\arraybackslash}p{1.2cm}|>{\centering\arraybackslash}p{1.2cm}|}
\hline
\textbf{Model} & \multicolumn{2}{c|}{\textbf{Original}} & \multicolumn{2}{c|}{\textbf{RealGuidance-Backtr.}} \\
\cline{2-5}
 & \textbf{Acc} & \textbf{F1-score} & \textbf{Acc} & \textbf{F1-score} \\
\hline
BERT-VGG & 0.8335 & 0.8277 & 0.8387 & 0.8389 \\
\hline
BERT-ViT & 0.8597 & 0.8588 & 0.8628 & 0.8621 \\
\hline
RoBERTa-ViT & 0.8702 & 0.8682 & 0.8785 & 0.8784 \\
\hline
RoBERTa-MambaViT & \textbf{0.8880} & \textbf{0.8870} & \textbf{0.8848} & \textbf{0.8831} \\
\hline
\end{tabular}

\end{table}

However, for the model that performed best on the original dataset, RoBERTa-MambaViT, this mix of augmentations resulted in a slight decrease in performance. This implies that the introduced variations may have interfered with the advanced feature extraction mechanisms of the architecture, highlighting the sensitivity of complex models to the noise caused by the augmentations.

The combination of DiffuseMix image augmentation with back-translated text also had mixed results (see Table~\ref{tab:tabel_10}). The RoBERTa-ViT model showed a slight performance improvement, likely due to the additional diversity. On the other hand, RoBERTa-MambaViT experienced a slight drop in performance, similar to the Real Guidance scenario. These results suggest that, while augmentations can be beneficial, their effectiveness depends on the underlying model architecture and its ability to augment multimodal information.

On the original CrisisMMD dataset, multimodal classification consistently outperformed unimodal text or image models, confirming the benefits of combining both modalities.

\begin{table}[ht]
\centering
\caption{Performance (Accuracy, F1-score) of multi-modal models with Diffusemix and Back-translation augmentations.}
\label{tab:tabel_10}
\renewcommand{\arraystretch}{1.25}
\setlength{\tabcolsep}{3pt}
\begin{tabular}{|>{\centering\arraybackslash}p{2.0cm}|>{\centering\arraybackslash}p{1.0cm}|>{\centering\arraybackslash}p{1.5cm}|>{\centering\arraybackslash}p{1.2cm}|>{\centering\arraybackslash}p{1.2cm}|}
\hline
\textbf{Model} & \multicolumn{2}{c|}{\textbf{Original}} & \multicolumn{2}{c|}{\textbf{Diffusemix-Backtr.}} \\
\cline{2-5}
 & \textbf{Acc} & \textbf{F1-score} & \textbf{Acc} & \textbf{F1-score} \\
\hline
RoBERTa-ViT & 0.8702 & 0.8682 & 0.8733 & 0.8738 \\
\hline
RoBERTa-MambaViT & \textbf{0.8880} & \textbf{0.8870} & \textbf{0.8827} & \textbf{0.8821} \\
\hline
\end{tabular}
\end{table}

\subsection{Results for Multi-View Learning}

Although the multi-view learning approach introduced greater model complexity and a higher capacity for capturing inter-modal dependencies, it did not outperform the baseline multimodal models (see Table~\ref{tab:tabel_11}). Specifically, attention-based multi-view models, which incorporated back-translation for text augmentation and Real Guidance for image augmentation, underperformed after 10 training epochs.

\begin{table}[ht]
\centering
\caption{Performance of multi-view learning.}
\label{tab:tabel_11}
\renewcommand{\arraystretch}{1.25}
\setlength{\tabcolsep}{3pt}
\begin{tabular}{|>{\centering\arraybackslash}p{2.8cm}|>{\centering\arraybackslash}p{1.8cm}|>{\centering\arraybackslash}p{1.8cm}|}
\hline
\textbf{Model}  & \textbf{Acc} & \textbf{F1-score} \\
\hline
RoBERTa-ViT & \textbf{0.8597} & \textbf{0.8607} \\
\hline
RoBERTa-ViT + cross-attention & 0.8461 & 0.8411 \\
\hline
RoBERTa-ViT + self-cross-attention & 0.8272 & 0.8228 \\
\hline
\end{tabular}
\end{table}

One possible explanation is that the increased parameters and complexity of the attention-based models require more extensive training to converge. Additionally, during the evaluation, only the original text-image pairs were provided, while the additional augmented views were left empty. This mismatch between the multi-view training strategy and the classic multimodal inference process may have prevented the model from fully benefiting from the additional features, limiting its performance.

These results suggest that while multi-view learning introduces promising architectural flexibility, its practical benefits depend on careful training and alignment between the training and evaluation processes.

\subsection{Comparison with Existing Work on CrisisMMD}
%In this section, the results of the proposed methods are compared with those reported in previous studies.

CrisisMMD has been widely adopted as a benchmark for disaster assessment, supporting the development of both unimodal and multimodal approaches. We note that not all the previous studies utilize the same variant of the CrisisMMD dataset. The experiments presented in this work were conducted using the agreed-upon annotation subset of CrisisMMD, focusing on the humanitarian classification task.

\subsubsection{Text Classification}

Table~\ref{tab:unimodal_text_results} presents the comparison of text-only classification results. The suggested RoBERTa-large model achieved a high F1-score, which is much better than the previous baseline set by Ofli et al.~\cite{ofli} and shows competitive performance relative to the more recent approach of Abavisani et al.~\cite{abavisani}.

\begin{table}[ht]
\centering
\caption{Comparison of unimodal text classification results on CrisisMMD.}
\renewcommand{\arraystretch}{1.2}
\begin{tabular}{|l|c|c|}
\hline
\textbf{Model} & \textbf{Accuracy} & \textbf{F1-score} \\
\hline
Ofli et al.\cite{ofli} & 0.7048 & 0.6770 \\
\hline
Abavisani et al. \cite{abavisani} & \textbf{0.8610} & \textbf{0.8780} \\
\hline
Pranesh  \cite{pranesh} & - & 0.7456 \\
\hline
RoBERTa-large (ours)  & 0.8470 & 0.8450 \\
\hline
\end{tabular}
\label{tab:unimodal_text_results}
\end{table}

\subsubsection{Image Classification}

Table~\ref{tab:unimodal_images_results} shows the comparison for image-only classification. The MambaViT model achieved an F1-score of 82.20\%, outperforming the earlier results of Ofli et al.~\cite{ofli} and Pranesh et al.~\cite{pranesh}, and approaching the accuracy reported by Abavisani et al.~\cite{abavisani}.

\begin{table}[ht]
\centering
\caption{Comparison of unimodal images classification results on CrisisMMD.}
\renewcommand{\arraystretch}{1.2}
\begin{tabular}{|l|c|c|}
\hline
\textbf{Model} & \textbf{Accuracy} & \textbf{F1-score} \\
\hline
Ofli et al.\cite{ofli} & 0.7680 & 0.7630 \\
\hline
Abavisani et al. \cite{abavisani} & \textbf{0.8340} & \textbf{0.8690} \\
\hline
Pranesh  \cite{pranesh} & - & 0.6835 \\
\hline
MambaViT (ours)  & 0.8240 & 0.8220 \\
\hline
\end{tabular}
\label{tab:unimodal_images_results}
\end{table}

\subsubsection{Multimodal Classification}

The comparison of the multimodal classification results is summarized in Table~\ref{tab:multimodal_results}. The proposed RoBERTa-MambaViT model achieved an F1-score of 88.7\%, demonstrating competitive performance among the state-of-the-art multimodal approaches. Although it did not surpass the best-performing score reported by Abavisani et al.~\cite{abavisani}, it significantly outperformed earlier baselines and approaches.

\begin{table}[ht]
\centering
\caption{Comparison of multimodal classification results on CrisisMMD.}
\renewcommand{\arraystretch}{1.2}
\begin{tabular}{|l|c|c|}
\hline
\textbf{Model} & \textbf{Accuracy} & \textbf{F1-score} \\
\hline
Ofli et al.\cite{ofli} & 0.7840 & 0.7830 \\
\hline
Abavisani et al. \cite{abavisani} & \textbf{0.9110} & \textbf{0.9180} \\
\hline
COBRA \cite{cobra} & 0.7240 & - \\
\hline
Pranesh  \cite{pranesh} & - & 0.8510 \\
\hline
Sirbu et al. \cite{anonym-crisis} & 0.8730 & 0.8720 \\
\hline

RoBERTa-MambaViT (ours) & 0.8880 & 0.8870 \\
\hline
\end{tabular}
\label{tab:multimodal_results}
\end{table}

%These results highlight that the proposed approaches deliver competitive performance across text, image, and multimodal classification tasks.

\section{Conclusions}

This paper applied text and image augmentation and evaluated their impact across unimodal, multimodal, and multi-view learning setups to improve disaster assessment on the CrisisMMD dataset.
Overall, we demonstrated the potential of diffusion-based augmentations for image data, and simple yet effective text augmentation techniques could improve disaster assessment models. The results also highlighted the challenges of effectively integrating multiple data sources, particularly when implementing complex learning strategies such as multi-view learning. Future work will focus on improved augmentation filtering and evaluation on more disaster-related datasets.

%\section*{Acknowledgments}
%This work is supported by the Romanian Hub for Artificial Intelligence - HRIA, Smart Growth, Digitization and Financial Instruments Program, 2021-2027, MySMIS no. 334906.

\bibliographystyle{IEEEtran}
\bibliography{mybib}

\end{document}